\documentclass[11pt]{article}

\usepackage[labelfont=bf,labelsep=period]{caption}
\usepackage{realA4}

\usepackage[utf8]{inputenc}
\usepackage{times}
\usepackage[pdftex]{hyperref}
\usepackage{color}
\usepackage[caption=false]{subfig}
\usepackage{multirow}
\usepackage{balance}
\usepackage{xspace}
\usepackage{float}
\usepackage{graphicx}
\usepackage{url}
\usepackage{booktabs}

\usepackage{mathtools}

\usepackage[printonlyused]{acronym}
\acrodef{SGX}[SGX]{Intel's Software Guard Extensions}
\acrodef{EPC}{Enclave Page Cache}
\acrodef{EPCM}{Enclave Page Cache Map}
\acrodef{MMU}{Memory Management Unit}
\acrodef{AEX}{Asynchronous Enclave Exit}
\acrodef{AEP}{Asynchronous Exit Pointer}
\acrodef{ISR}{Interrupt Service Routine}
\acrodef{OS}{operating system}
\acrodef{DoS}{denial-of-service}
\acrodef{SDK}{Software Development Kit}
\acrodef{TCB}{trusted computing base}
\acrodef{EDL}{Enclave Description Language}
\acrodef{API}{application programming interface}
\acrodef{TEE}{Trusted Execution Environment}
\acrodefplural{TEE}{trusted execution environments}

\usepackage{todonotes}

\usepackage{amsmath} 
\allowdisplaybreaks[2]          

\usepackage{amssymb} 

\usepackage{amsthm} 
\newtheoremstyle{plain-boldhead}
  {\topsep}
  {\topsep}
  {\itshape}
  {}
  {\bfseries}
  {.}
  { }
  {\thmname{#1}\thmnumber{ #2}\thmnote{ (\bfseries #3)}}
\newtheoremstyle{definition-boldhead}
  {\topsep}
  {\topsep}
  {\normalfont}
  {}
  {\bfseries}
  {.}
  { }
  {\thmname{#1}\thmnumber{ #2}\thmnote{ (\bfseries #3)}}
\theoremstyle{plain-boldhead}

\theoremstyle{definition-boldhead}
\newtheorem{definition}{Definition}

\floatstyle{ruled}
\newfloat{algo}{htb}{alg}
\floatname{algo}{Algorithm}

\newcommand{\N}{\mathbb{N}}

\newcommand{\false}{\str{false}\xspace}

\newcommand{\str}[1]{\textsc{#1}}
\newcommand{\var}[1]{\textit{#1}}
\newcommand{\op}[1]{\textsl{#1}}
\newcommand{\msg}[2]{\ensuremath{[\str{#1}, {#2}]}}

\newcommand{\tup}[1]{%
  \relax\ifmmode%
    \langle#1 \rangle%
  \else
    $\langle$#1$\rangle$%
  \fi}

\newcommand{\CC}{\ensuremath{\mathcal{C}}\xspace}

\newcommand{\CK}{\ensuremath{\mathcal{K}}\xspace}

\newcommand{\CS}{\ensuremath{\mathcal{S}}\xspace}

\newcommand{\ASSERT}{\textbf{assert}\xspace}

\newcommand{\DO}{\textbf{do}\xspace}
\newcommand{\ELSE}{\textbf{else}\xspace}

\newcommand{\FUNCTION}{\textbf{function}\xspace}

\newcommand{\IF}{\textbf{if}\xspace}

\newcommand{\RETURN}{\textbf{return}\xspace}
\newcommand{\STATE}{\textbf{state}\xspace}

\newcommand{\UPON}{\textbf{upon}\xspace}

\newcommand{\pproj}{Lightweight Collective Memory\xspace}
\newcommand{\pp}{LCM\xspace}

\hypersetup{
    pdfsubject = {},
    pdftitle = {\pproj},
    pdfauthor = {},
    pdfkeywords = {Cloud Computing},
    pdfborder = {0 0 .5}
}

\begin{document}

\title{\bf Rollback and Forking Detection for Trusted Execution Environments using Lightweight Collective Memory}

\author{
  Marcus Brandenburger\\
  IBM Research - Zurich\\
  \url{bur@zurich.ibm.com}
  \and Christian Cachin\\
  IBM Research - Zurich\\
  \url{cca@zurich.ibm.com}\\
  \and  Matthias Lorenz\\
  TU Braunschweig\\
  \url{mlorenz@ibr.cs.tu-bs.de}\\
  \and  R\"udiger Kapitza\\
  TU Braunschweig\\
  \url{rrkapitz@ibr.cs.tu-bs.de}
}

\date{\today}

\maketitle

\begin{abstract}
Novel hardware-aided trusted execution environments, as provided by \ac{SGX},
enable to execute applications in a secure context that enforces
confidentiality and integrity of the application state even when the host
system is misbehaving.
While this paves the way towards secure and trustworthy cloud
computing, essential system support to protect persistent application state 
against rollback and forking attacks is missing.

In this paper we present \emph{LCM} -- a lightweight protocol to establish a
collective memory amongst all clients of a remote application to detect
integrity and consistency violations.  
LCM enables the \emph{detection of rollback attacks} against the remote
application, enforces the consistency notion of
\emph{fork-linearizability} and notifies clients about operation
stability.  The protocol exploits the trusted execution environment,
complements it with simple client-side operations, and maintains only
small, constant storage at the clients.
This simplifies the solution compared to previous approaches, where
the clients had to
verify all operations initiated by other clients.
We have implemented LCM and demonstrated its advantages with a key-value store
application.  The evaluation shows that it introduces low network and
computation overhead; in particular, a LCM-protected key-value store 
achieves 0.72x -- 0.98x of a SGX-secured key-value store throughput.
\end{abstract}

\acresetall

\section{Introduction}\label{sec:intro}

Despite numerous efforts by industry and academia cloud computing suffers still from trust issues~\cite{Jayaram2014,Pearson2010}.
This is not surprising as companies possess limited control once their applications and data enter the cloud. 
Users have to trust the operating personal and a complex software stack composed of management software, virtualization layers, as well as commodity operating systems. 
On top, cloud providers are typically reluctant to share their exact system details because this information is critical for their business. 

The recently released Software Guard Extensions (SGX)~\cite{mckeen2013} technology of Intel is expected to make a change, as it addresses trust issues that customer face when outsourcing services to off-site locations and still gives cloud providers the freedom to not disclose their system details.
SGX offers an instruction set extension that allows to establish trusted execution contexts, called \emph{enclaves}.
These enclaves might be tailored and comprise only a small dedicated fraction of an application~\cite{hoekstra2013usinginnovative,mw16seckeeper} or can contain an entire legacy application and the
necessary operating system support~\cite{baumann2014,scone-osdi16}.
Thereby, the plaintext of enclave-protected data and code is only available for computation inside the CPU, and encrypted as soon as it leaves the CPU package again. 
In this way, enclave-residing data is even guarded against unauthorized accesses by higher privileged code and from attackers with administrative rights and physical access.

While SGX can be considered as a big step forward towards trustworthy cloud computing, some attack vectors nevertheless remain. 
One important open issue are \emph{rollback} and \emph{forking attacks} on stateful applications that make use of persistent storage. 
Whereas SGX provides mechanisms against main-memory replay attacks, persistent storage is not under the direct control of SGX and therefore harder to secure.
The need to handle system restarts, operating system crashes, and power outages makes a completely secure solution for state continuity difficult to achieve.
Baumann et al.~\cite{baumann2014} who pioneered the field by proposing \emph{application enclaves} acknowledge this issue and suggest to use a central external service that is contacted on every request. 
However, this only delegates the problem to an external entity, demands additional remote communication and adds another single point of failure.
Strackx and Piessens~\cite{strpie16}, on the other hand, proposed abstractions on top of hardware-based trusted counters. 
This and similar approaches \cite{chun2007,levin2009,parno2011,strpie16} enable immediate detection of forking attacks but suffer from bad performance, as writing and reading trusted non-volatile counters for every request is time-consuming.
Finally, there are a number of approaches that do not rely on secure execution contexts, such as enclaves, but utilize only plain resources of an untrusted provider~\cite{brcakn15,cakesh11,cacohr14,fzff10,mazsha02,scckms10}.
These systems typically require cooperating clients to verify each server
response.  In particular, this comes with additional communication
overhead between clients and server, and requires costly
cryptographic verification.

In this paper we present \emph{\pproj (\pp)} -- a distributed protocol to establish a collective memory amongst all clients of a remote application to detect integrity and consistency violations. 
By leveraging  \acp{TEE}, such as SGX, \pp keeps client interaction and service state confidential. 
It ensures fork-linearizability~\cite{mazsha02}, which denotes the strongest consistency notion among the clients that can be achieved in the presence of rollback attacks without direct client-to-client communication and in absence of trusted non-volatile memory. 
Furthermore, \pp notifies clients about operation stability. 
This criteria refers to \emph{stable} operations where a client can be sure that its request has been acknowledged by a designated number of other clients. 
A typical size would be the majority of clients.
Finally, compared to previous approaches that rely on trusted counters,
applications secured by \pp can be migrated across physical TEEs
and maintain their capability
to detect rollback attacks and to enforce fork-linearizability.

We implemented \pp as a Java and C++ framework and demonstrate its advantage by
securing a key-value store.  We evaluated the performance of the prototype by
using the YCSB benchmark and compare with native execution and SGX-secured approaches.
It turns out that a {SGX}-secured key-value store achieves 0.42x -- 0.78x performance of
unprotected native execution.  However, the performance of \pp is 0.72x -- 0.98x of the SGX-secured key-value store throughput, while on top enabling rollback and forking
detection.

The remainder of the paper is structured as follows.
Sec.~\ref{sec:model} provides a detailed problem description
and outlines the necessary background.  In Sec.~\ref{sec:forking},
past solutions for state continuity are discussed and the goals of
\pp are stated.  
Next, the overall architecture and the \pp protocol are introduced
in Sec.~\ref{sec:protocol}.
Sec.~\ref{sec:impl} provides the details of our implementation using SGX.
Subsequently, Sec.~\ref{sec:eval} explains the evaluation results.
Finally, Sec.~\ref{sec:related} outlines related approaches while Sec.~\ref{sec:conclusion} concludes the paper.

\section{Problem description}\label{sec:model}

\subsection{System model}

We consider an asynchronous distributed system with~$n$
\emph{clients}~$C_1,...,C_n$ and a \emph{server}~$S$.  The server contains
a \emph{trusted execution environment~(TEE)}, which hosts a \emph{trusted
  execution context}~$T$; this is an isolated, protected container that
runs an application protocol and is trusted by the clients.
A protocol~$P$ specifies the behavior of the clients, the server~$S$, and
the trusted execution context~$T$.  All clients are \emph{correct},
follow~$P$, and mutually trust each other; clients and the server may crash
but are able to recover with the help of stable storage, which they can
access through \op{load} and \op{store} operations.
In contrast,~$T$ is correct but runs under the control of~$S$ as explained
in detail later; $T$ does not have direct access to stable storage and may
lose its state.
The server is either correct and follows~$P$ or is \emph{Byzantine},
deviating arbitrarily from~$P$.  

\begin{figure}[ht]
    \centering
   \includegraphics[width=12cm]{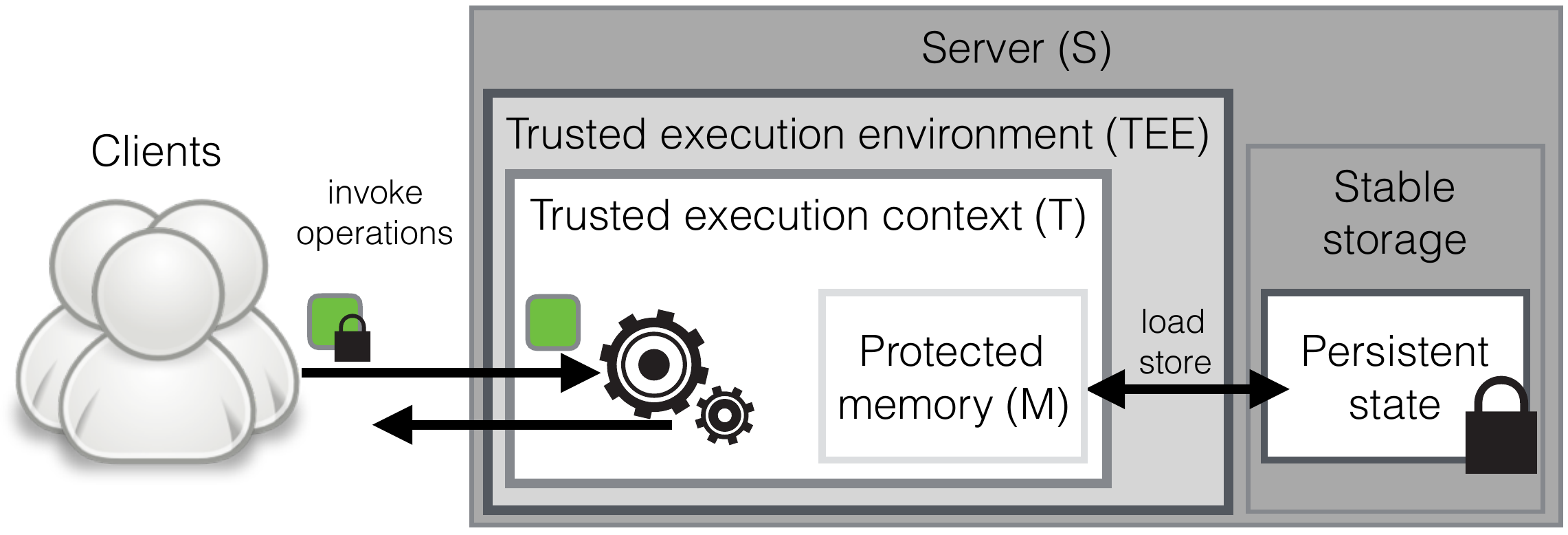}
    \caption{
    System model comprising trusted clients, a potentially misbehaving server
    $S$ that hosts a trusted execution context~$T$.}
    \label{fig:model}
\end{figure}

The clients and $T$ interact by exchanging
messages as specified by~$P$.  They communicate indirectly through the server
which should forward messages among them.  If $S$ is correct, then their
communication is reliable and respects \emph {first-in first-out (FIFO)} 
semantics; otherwise, $S$ may arbitrarily interfere with their messages.
Clients have limited communication capabilities beyond this and
do not interact with each other normally. 
The clients invoke a stateful application functionality~$F$, which provides
a set of operations; $F$ defines a response and a state
change for every operation.  The operations are executed by~$T$ inside the
TEE and, therefore, the state of $F$ is protected from a potentially
malicious~$S$.
We use the standard notions of executions, histories, sequential histories,
real-time order, concurrency, and well-formed executions from the
distributed-computing literature~\cite{attwel04}.  In particular, every operation execution is
represented by an \emph{invocation event} and a \emph{response event}.  An
operation is called \emph{complete} when a client receives a response
event.  Two operations are \emph{concurrent} if the invocation event of one
of them occurs before the other operation is complete.

\subsection{Trusted execution context}

A TEE provides a secure context for executing applications, isolated from
the server that hosts the TEE.  It protects the confidentiality and
integrity of code and data for the application running inside the execution
context.  More specifically,
a trusted execution context~$T$ is instantiated with a protocol~$P$, which
defines the program code executed by~$T$. After server~$S$ has created some
$T$, $S$ may start, terminate, and restart $T$ at
its discretion.
Once~$T$ has been created, $P$ running within $T$ cannot be modified
anymore nor may any other protocol~$P'$ be executed in~$T$.
The server may also create and run multiple
instances of~$T$ concurrently.
The time between instantiation and termination of~$T$ is called an
\emph{epoch}.  The entire lifetime of a trusted execution context can span
multiple epochs.

The TEE provides access to a secure random number generator that allows to
build cryptographic primitives, such as key generation, encryption and digital
signatures.
The TEE operates a cryptographic key-management infrastructure rooted
in a secret key protected by the TEE, which may
provide a program-specific key to a trusted execution context.
That is, a function $\op{get-key}_{T,P}$ is available to $T$ when
it executes protocol~$P$ and returns a secret key~$k$ that is specific
to~$P$ and the TEE.  Another $T'$, which is also instantiated with $P$, obtains the
same $k$, but any $T$ running~$P' \neq P$ or any other TEE 
obtains a key different from~$k$.

The clients can verify that a trusted execution context has been
instantiated with a certain protocol~$P$ and that~$P$ is indeed running
inside the TEE. This is essential for the assumption that~$T$ is trusted.
For this purpose clients leverage a procedure called remote
attestation~\cite{anati2013}.
In short, a client with prior information about $P$ sends a challenge to
$T$ and in return receives a cryptographic proof~$\phi$ that reflects~$P$
and the underlying TEE.  The client then verifies~$\phi$ and becomes
convinced that~$T$ runs~$P$, based on the cryptographic protocol and on its
trust in the~TEE.

Furthermore, $T$ is equipped with a small protected memory area~$M$ that
can only be accessed by~$T$.  It holds the execution-specific state as
defined by~$P$. Neither the server nor any other trusted execution context
can access or modify~$M$.
The protected memory is volatile, thus $M$ is only accessible within an
epoch of~$T$. In other words, when~$T$ stops, crashes, or restarts, then $M$
is lost.  This is not an issue for stateless protocols, but services
without state are generally not very useful; in realistic applications,
where the server maintains some state, $M$ must be restored after~$T$ has
been restarted.
For this reason~$M$ is stored externally on stable storage using \op{load}
and \op{store}, so that $T$ can access state from another epoch.

\subsection{Threats}

Normally server $S$ is correct, but it may become \emph{malicious} and
behave incorrectly, when corrupted by an attacker or affected by a software
bug.  A malicious server has full control over the
operating system, applications, and the data residing in memory and
stable storage,
but it cannot tamper with code and data in the trusted execution context.
This means $T$ is correct and follows~$P$ even though $S$ is malicious.

However, $S$ controls every interaction of $T$ with the environment.  A
malicious server may intercept, modify, reorder, discard, or replay
messages to and from~$T$.  Although some of those attacks can be prevented
by establishing a secure channel between a client and~$T$, a malicious $S$
may simply discard their messages; such a \emph{denial-of-service (DoS)}
attack is outside the scope of this work, however.

The trusted execution context must consider anything that it receives as
untrusted.  In particular, this holds when~$T$ accesses the stable storage
through \op{load} and \op{store}, in order to persist its state~$M$.  With
a correct $S$, \op{load} always returns the state that has been
\op{stored} most recently.  For protecting against a malicious~$S$, the
trusted execution context uses encryption and authentication to protect~$M$
before it leaves~$T$.
Yet, a malicious server may still return a correctly protected but outdated
state to~$T$.  We call such a consistency violation a \emph{rollback
  attack}.  In particular, a malicious server may restart~$T$ at any time
and \op{load} its memory from some state that~$T$ has \op{stored} earlier.

Furthermore, a malicious server may start multiple instances of a trusted
execution context and let the clients interact with different instances over
time. In this way, clients may be separated so that they only see
operations of other clients talking to the same instance.  
Even if the TEE can run only a single~$T$ at a time, $S$ can multiplex 
different copies of the trusted context. 
The malicious server might supply a different, but valid state to each
trusted execution context instance, similar to a rollback attack.  This
clearly violates the consistency of the data, so that the responses from
different trusted execution contexts to the clients diverge.  We call this
a \emph{forking attack}; it is more general than a rollback attack because
multiple instances of~$T$ answer concurrently to the clients.
Note that with a single instance of $T$ a forking attack always involves at
least one rollback attack.
It is well-known that clients cannot detect rollback and
forking attacks in asynchronous systems, unless they communicate directly
with each other~\cite{mazsha02}.

\section{Protecting against Forking Attacks}
\label{sec:forking}

\subsection{Trusted monotonic counters}
\label{subsec:tmc}

For defending an execution context~$T$ against a forking attack, we need to
assure \emph{state continuity}, i.e., that the state of~$T$ evolves
continuously and is never rolled back.  One might think that $T$ could simply 
maintain a cryptographic hash of~$M$ inside the TEE whenever it
\op{stores}~$M$ and verify that upon a \op{load} operation.  However, this
does not work because the memory of $T$ and the TEE is volatile and
disappears when the epoch ends.

To overcome this,~$T$ will need non-volatile storage that survives reboots.
Such defenses have been proposed in the form of an \emph{attested
  append-only memory} (A2M)~\cite{cmsk07} or a \emph{trusted incrementor}
(TrInc)~\cite{levin2009}.  These works demonstrate that the functionality
needed from the trusted non-volatile storage can be reduced to a
\emph{trusted monotonic counter}~(TMC).

In more detail, suppose $T$ has access to a TMC that is located in the TEE,
the TMC uses a non-volatile storage location that survives power loss, and
the TMC's state and its communication with~$T$ are protected from~$S$.
Whenever~$T$ \op{stores} $M$ at the untrusted server, it increments the
counter and includes the counter value with the state. When~$T$ is
restored, e.g. after a reboot, it \op{loads} its state from $S$, extracts
the counter value, reads the TMC, and compares it to the extracted counter.
Since~$T$ protects all \op{stored} data cryptographically with a key known
only inside the TEE, the server cannot tamper with the counter attached
to~$M$. This allows $T$ to detect rollback attacks.
However, this approach suffers from several disadvantages as we argue now.
 
First, it is not easy to tolerate concurrent crashes and maintain
liveness~\cite{parno2011} at the same time; that is, when~$T$ has incremented the TMC but
the server crashes before the counter value has been saved to the
non-volatile trusted area, then~$T$ might falsely accuse the correct server
of performing a rollback attack.  The reason is $T$ cannot differentiate
between a rollback attack and a server crash.
In order to tolerate crashes, one can resort to complex schemes that ensure
state continuity, which increment the TMC, save it persistently, and write
state to disk atomically; they either need
hardware modifactions~\cite{strackx2014} or perform a variation of 2-phase
commit~\cite{parno2011,strpie16}, but the latter only works for deterministic
operations, which can be replayed by $T$ and always give the same output.

Second, TMC-based solutions often suffer from limited performance.
Typically, TMCs are implemented using TPMs which are well known to be
slow~\cite{parno2011}.  The reason is that in order to prevent a counter
overflow, the TMC artificially increases the time to increment the counter
to several milliseconds.  Although a response time of a several
milliseconds is acceptable for, say, digital right management (DRM), this
has a negative impact on the throughput of a server application
that processes requests at a high rate.

Finally, the main disadvantage of any TMC-based approach is the lack of
location transparency.  That is, the TMC is normally bound to one trusted
execution environment within one server. However, in modern cloud-com\-pu\-ting
platforms, applications must be able to scale and run on different servers
during their lifetime.  This may already be caused by system
maintenance.  For the end-user this should be completely transparent, but a
trusted execution context cannot be stopped on one server and restarted on
a different server with the same TMC; this is exactly what trusted
hardware should prevent.  Therefore this would require a migration
protocol that needs the help of a trusted party.

For these reasons, we do not consider any solution that requires extra
hardware or restricts the application to be deterministic in this paper.
Instead we exploit the guarantees available with standard TEEs.

\subsection{Ensuring consistency at the clients}

In the model considered here the TEE does not prevent a malicious server
from mounting rollback and forking attacks and from isolating the clients
from each other.  The best possible option is to ensure that the clients
remain ``synchronized'' with each other as much as possible and to mitigate
attacks through this.

\subsubsection{Fork-linearizability}

\emph{Fork-linearizability}~\cite{mazsha02} denotes the strongest consistency
notion among the clients that can be achieved in the presence of rollback attacks
and without client-to-client communication.  This well-established notion
ensures that whenever the malicious server has separated two clients, they
can never be joined again to see mutually inconsistent responses from the
server, without one of them detecting the attack.  In essence, the server
has to pretend that the inconsistency remains forever.  Clearly, the
clients can detect this though a lightweight, out-of-band mechanism.

Protocols that ensure fork-linearizability work by embedding information
about the causal past of each operation into the requests from client to
servers~\cite{mazsha02,cacohr14,brcakn15}.  They use hash chains, Merkle
trees, and vector clocks for representing the past history of operations
and their context.  Such protocols are very similar to the use of hash
chains in blockchain platforms~\cite{bmcnkf15}, cryptocurrencies such as
Bitcoin, and Certificate Transparency~\cite{laurie14}.

The standard notion of \emph{linearizability}~\cite{herwin90} requires that
the operations of all clients appear to execute atomically in one sequence,
and that the atomic sequence respects the real-time partial order of the
operations that the clients observe.  \emph{Fork-linearizability} is
defined as an extension of this, which relaxes the condition of one
sequence to permit multiple ``forks'' of an
execution~\cite{mazsha02,cashsh07}.  Under fork-linearizability, every
client observes a linearizable history and when an operation is observed by
multiple clients, the history of events occurring before the operation is
the same.  In this context, the \emph{view} of a client~$C_i$ denotes a
correct, serialized history of operations for the functionality~$F$, which
includes all operations of~$C_i$.  For a more formal treatment we refer to
the literature~\cite{cashsh07}.

Unfortunately, fork-linearizability cannot be achieved without taking into
account that some client operations on a correct server are blocked until
other, concurrent operations terminate~\cite{cashsh07}.  This inherent
limitation has led to the relaxed notions, such as weak
fork-linearizability.  In FAUST~\cite{cakesh11}, for instance, an operation
returns a response to the client that is not guaranteed to be immediately
fork-linearizable or linearizable, but the protocol notifies the client
later when it knows that other clients have observed the operation as well.
This is captured by the notion of \emph{stability}, discussed next.

\subsubsection{Operation stability}

We now define a way to inform the client about those of its
operations that  have reached some level of consistency
with respect to other clients.
More precisely, we call an operation~$o$ by a client~$C_i$ \emph{stable}
with respect to another client~$C_j$ if the views of $C_i$ and $C_j$ both
include~$o$.  In other words, $C_i$ knows that $C_j$ has observed~$o$
and that $S$ was forced to take into account any effects of~$o$ in later
service responses to~$C_j$.  

Operation stability has also been used by~\cite{ttpdsh95, cakesh11}.
Here we use it as follows.  We augment the response event of every
operation with two numbers: a \emph{sequence number}, which is
assigned by the protocol to the operation that completes; and a
\emph{stable sequence number}, which denotes the latest stable
sequence number of this client.  The sequence numbers returned at one client are strictly increasing; the stable
sequence numbers never decrease.

\begin{definition}[Operation stability]
  \label{def:op-stability}
  Let $o$ be a complete operation of $C_i$ that returns sequence
  number~$t$. We say that $o$ is \emph{stable w.r.t.~a client $C_j \neq C_i$} after
  $C_j$ completes any operation that returns a sequence number that
  is bigger than~$t$.  Operation $o$ of $C_i$ is always stable w.r.t. $C_i$.

  For a set of clients~$G$ that includes~$C_i$, an operation $o$ of~$C_i$
  is \emph{stable w.r.t. the set of clients $G$}, when $o$ is stable
  w.r.t.~all~$C_j \in G$.  An operation that is stable w.r.t.~all clients
  is simply called \emph{stable}.
\end{definition}

One may use different strengths of stability; for example, an operation
might take a long time until it becomes stable (because all clients
must observe it), but it might already be stable at a subset of the clients
much earlier.  A particularly useful subset is a majority quorum of the
clients.

\begin{definition}[Operation stability among a maj\-ori\-ty~\cite{scckms10}]
\label{def:op-stability-majority}
  An operation $o$ of $C_i$ is \emph{stable among a majority} of clients, when
  $o$ is stable w.r.t. a set of clients $C$, where $|C| > n/2$.
\end{definition}

Note that any subsequence of a history that contains only operations that are stable among a majority is linearizable.

\section{\pproj}\label{sec:arch}
\label{sec:protocol}

This section introduces \emph{\pproj} \emph{(\pp)}, a protocol that allows
a group of mutually trusting clients to run a service on a (potentially
malicious) remote server.  It benefits from a trusted execution context~$T$
that runs on the server and executes the operations on behalf of the
clients.
\pp facilitates the detection of forking and rollback attacks against $T$
by ensuring fork-linearizability for every client operation.  Moreover, \pp
indicates which operations are stable among a majority; this permits
clients to infer when their operations are linearizable.
The \pp protocol benefits from the security guarantees of the TEE; in
contrast to all previous protocols in the line of work originating with
Mazi{\`e}res and Shasha~\cite{mazsha02}, aiming at fork-linearizable
semantics, the clients do not verify operation results here.  Clients only
handle metadata and rely on the TEE for producing correct responses.

\subsection{Overview}

\pp executes a stateful functionality~$F$ inside a trusted execution
context~$T$ that is instantiated with the \pp protocol and also runs~$F$.
The trusted~$T$ constructs a hash chain from the history of all operations
that it executes and embeds this information in its responses to the
clients.

A client invokes an operation by sending an encrypted $\str{invoke}$
message to the (untrusted) server~$S$, which forwards all incoming messages
to~$T$.
After $T$ has decrypted this, it first verifies that the view of the client
is consistent with~$T$'s own history.  Then $T$ executes the operation and
assigns a sequence number to it.  The operation produces an output for the
client and may modify the state of~$F$.  The output is returned to the
client in a $\str{reply}$ message, together with the sequence number and
the latest stable operation (represented as a sequence number).
When the client receives the $\str{reply}$ message, it completes the
operation and returns the result, the assigned sequence number, and the
majority-stable sequence number.  The latter informs the client about the
stability of its earlier operations.

\begin{figure}[ht]
    \centering
   \includegraphics[width=12cm]{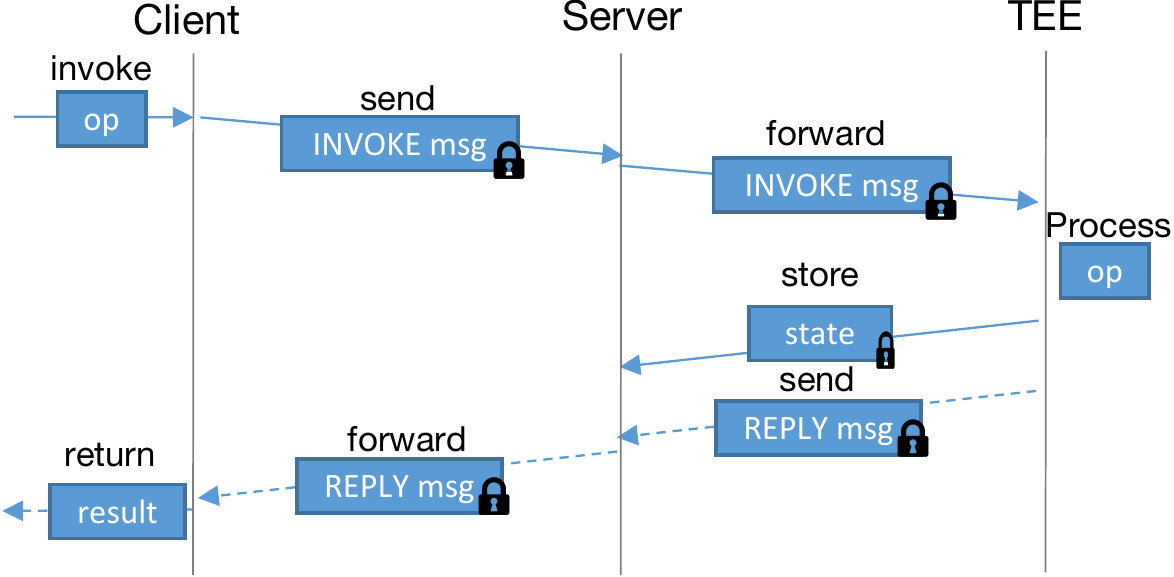}
    \caption{Protocol messages in \pproj}
    \label{fig:protocl_flow}
\end{figure}

Fig.~\ref{fig:protocl_flow} shows the protocol interaction.
For simplicity we assume that each client invokes operations sequentially,
that is, it invokes a new operation only after completing the
previous operation.
For protecting~$T$ against a malicious $S$, three cryptographic keys are
used:
\begin{enumerate}
\item To safeguard the protocol's consistency data in the hash chain and
  the service state, $T$ encrypts it with a \emph{protocol-state encryption
    key}~$k_{P}$ before \op{storing} it in the server's stable storage and
  decrypts it again after a \op{load}.  This key is generated by an admin
  during bootstrapping and required for migrating~$T$ to another server.

\item A \emph{sealing key}~$k_S$ is initially generated by the TEE using
  $\op{get-key}_{T,\pp}$ when $T$ starts.  It encrypts the protocol-state
  key~$k_P$ when $T$ \op{stores} this in persistent storage to tolerate
  crashes.

\item A \emph{communication key}~$k_{\CC}$ protects all messages exchanged
  between the clients and~$T$.  The key is also generated by an admin and
  made known to all clients and to~$T$.
\end{enumerate}

All encryption operations use \emph{authenticated encryption} with a
symmetric-key~$k$ and two functions $\op{auth-encrypt}(m, k)$ and
$\op{auth-decrypt}(c, k)$ for a message~$m$ and ciphertext~$c$.
Authenticated encryption produces a ciphertext integrated with a
message-authentication code~(MAC); it protects the content from leaking
information to~$S$ and prevents that $S$ tampers with messages or stored
data by altering ciphertext.
The \emph{hash function} in \pp, denoted $\op{hash}()$, can be any
cryptographically secure collision-free hash function; it maps a bit
string~$x$ of arbitrary length to a short, unique hash value~$h$.

\subsection{Protocol}
\label{sec:proto-details}

\subsubsection{Invocation at the client (Alg.~\ref{alg:client})}

The client uses variables $t_c$ and~$t_s$ to hold sequence
numbers for the last operation by $C_i$ and the last operation stable among
a majority, respectively.  In addition, the client stores $h_c$, the hash
chain value computed by~$T$ corresponding to its most recent operation
(with sequence number~$t_c$).
When $C_i$ invokes an operation $o$, it buffers~$o$ in a variable~$u$ and
sends an encrypted $\str{invoke}$ message containing $i$, $o$, $t_c$,
and~$h_c$.  The latter two values represent the context in which~$C_i$
invokes $o$; they result from $C_i$'s last operation.

\subsubsection{Execution at~$T$ (Alg.~\ref{alg:enclave})}

The trusted execution context~$T$ maintains the sequence number of the most
recently executed operation in a counter $t$ and a corresponding hash-chain
value in~$h$.
$T$ processes the operations of the clients sequentially. When~$T$ receives an
$\str{invoke}$ message from $C_i$, it decrypts the message with~$k_{\CC}$ and
signals a violation if the message does not have valid authentication.  Then
$T$ verifies that $(t_c, h_c)$ sent by the client correspond to the last
operation response that $T$ has returned to $C_i$.  For this purpose, $T$
maintains a map~$V$ indexed by client identifier, where entry~$V[i]$ holds the
sequence number of the last acknowledged operation by $C_i$, the
sequence number and corresponding hash chain value after the last operation
by~$C_i$. Again, when an inconsistency is detected, then $T$ halts.
This verification is essential for the protocol and has three goals:
First, it acknowledges the previous operation by $C_i$ in the sense that
$T$ learns that $C_i$ has actually received the reply for its last
invocation.
Second, this detects message-replay attacks.  When a malicious~$S$
forwards the same \str{invoke} message multiple times, $T$ can easily
filter these out with~$V$.
Finally, the verification detects rollback or forking attacks
because the client sends the condensed view of its own history contained in
$t_c$ and $h_c$.

If sequence number and hash chain value verification is successful, then $T$
increments the sequence number~$t$, and calls~$\op{exec}_F$, which applies the
operation~$o$ to state~$s$ and yields the corresponding result~$r$ according
to~$F$. Next, $T$ extends the hash chain $h$ by setting this to $\op{hash}(h
\| o \| t \| i)$.
With the information from the \str{invoke} message, $T$ also determines if
more operations have become stable.  It uses the data in $V$ and a function
\op{majority-stable} that returns~$q$, the highest sequence number of an
operation stable among a majority.

Then $T$ sends a $\str{reply}$ message to $C_i$ encrypted with $k_{\CC}$,
containing the sequence number~$t$, the hash chain value~$h$, the
result~$r$, the stable operation~$q$, and the client's previous hash chain
value $h_c$.  Before sending $\str{reply}$, $T$ also needs to \op{store}
the current state for recovering from a crash.  For this, $T$ encrypts the
service state~$s$, the protocol state~$V$, and the key~$k_{\CC}$ using
$\op{auth-encrypt}$ with $k_{P}$ and \op{stores} this as a \var{blob}
through~$S$.

\subsubsection{Verification at the client}

When $C_i$ receives a~$\str{reply}$ message, it uses $k_{\CC}$ to decrypt
the contents and extracts $t$, $h$, $r$, $q$, and~$h_c'$.  The client
verifies that the previous hash chain value~$h_c'$ is equal to its own $h_c$, in
order to match the~$\str{reply}$ message to its most recent $\str{invoke}$.
Next, $C_i$ stores the new sequence number and hash chain value $(t, h)$
and outputs the operation result~$r$ and the majority-stable operation~$q$.
These two sequence numbers allow the client to keep track of the operation
history.  In particular, a majority of clients have observed all operations
with sequence numbers up to~$q$.  Any operation of $C_i$ with the sequence
number~$t'\leq q$ is now stable among a majority.
For correct functioning of the protocol, the state of each client must be
recoverable from stable storage if a client crashes.  For simplicity this
is not part of the pseudocode.

\subsubsection{Server}

The (correct) server~$S$ runs a TEE and hosts~$T$, which is initially
created by an admin.  Whenever~$S$ reboots or crashes, it restarts~$T$.
Recall that a malicious~$S$ may restart the trusted execution context at
any time or even spawn multiple instances.
Furthermore, a correct $S$ forwards all messages between the clients and
the trusted execution context in FIFO order.  A malicious server, in
contrast, can discard, reorder or delay messages.

\subsubsection{Protocol details}
In the pseudocode in Alg.~\ref{alg:client}--\ref{alg:enclave}, the
symbol $\|$ denotes the concatenation of bit strings, and the
\textbf{assert} statement, parameterized by a condition (where $\ast$ matches
any value), immediately terminates the protocol when the condition is false.  The clients and $T$
use this to signal that the server misbehaved.  Note that \op{auth-decrypt}
may also signal an error; this is equivalent to an assert~\false statement.

\begin{algo}
\vbox{
\small
\begin{tabbing}
xxxx\=xxxx\=xxxx\=xxxx\=xxxx\=xxxx\=xxxx\kill
\STATE \\
\> $t_c \in \N_0$: last sequence number, initially $0$ \\
\> $t_s \in \N_0$: last majority-stable sequence number, initially $0$ \\
\> $h_c \in \{0,1\}^*$: last hash chain value, initially $h_c = h_0$ \\
\> $k_{\CC} \in \CK$: protocol key \\
\\
\FUNCTION $\op{invoke}(o)$ \\
\> $\var{invoke} \gets \op{auth-encrypt}(\msg{invoke}{t_c, h_c, o, i}, k_{\CC})$ \\
\> send message $\var{invoke}$ to $S$ \` \\
\\
\UPON receiving message $\var{reply}$ from $S$ \DO \\
\> $\msg{reply}{t, h, r, q, h_c'} \gets \op{auth-decrypt}(\var{reply}, k_{\CC})$ \\
\> \ASSERT $h_c' = h_c$ \\
\> $(t_c, t_s, h_c) \gets (t, q, h)$ \\
\> \RETURN $(r,t,q)$ \` // response of operation
\end{tabbing}
}
\caption{\pp Protocol for client~$C_i$}
\label{alg:client}
\end{algo}

\begin{algo}
\vbox{
\small
\begin{tabbing}
xxxx\=xxxx\=xxxx\=xxxx\=xxxx\=xxxx\=xxxx\kill
\STATE \\
\> $t \in \N_0$: sequence number, initially $0$ \\
\> $h \in \{0,1\}^*$: last hash chain value, initially $h = \bot$ \\
\> $V: \N \to \N_0 \times \N_0 \times \{0,1\}^*$: current protocol state, init. $[0]^N$ \\
\> $s \in \CS$: state of the service, initially $s = s_0$ \\
\> $k_{S} \in \CK$: sealing key, initially $k_{S} = \bot$ \\
\> $k_{P} \in \CK$: state encryption key, initially $k_{P} = \bot$ \\
\> $k_{\CC} \in \CK$: communication encryption key, initially $k_{\CC} = \bot$ \\
\\
\FUNCTION $\op{init}$ \\
\> $k_{S} \gets \op{get-key}_{T,P}$ \` // get sealing key \\
\> $(\var{blob}_\var{key}, \var{blob}_\var{state}) \gets \op{load}$ \` // possible rollback attack \\
\> \IF $\var{blob}_\var{key} = \bot$ \\
\> \> perform bootstrapping as described in the text\\
\> \ELSE \\
\> \> $k_{P} \gets \op{auth-decrypt}(\var{blob}_\var{key}, k_{S})$ \\
\> \> $(s, V, k_{\CC}) \gets \op{auth-decrypt}(\var{blob}_\var{state}, k_{P})$ \\
\> \> $(\cdot, t,h) \gets V[ \, \op{argmax}(V) \, ]$ \\
\\

\UPON receiving message $\var{invoke}$ from $C_i$ \DO \\
\> $\msg{invoke}{t_c, h_c, o, i} \gets \op{auth-decrypt}(\var{invoke}, k_{\CC})$ \\
\> \ASSERT $V[i] = (\ast, t_c, h_c)$ \\
\> $t \gets t + 1$ \\
\> $(r, s) \gets \op{exec}_F(s, o)$ \\

\> $h \gets \op{hash}(h \| o \| t \| i )$ \\
\> $V[i] \gets (t_c,t,h)$ \\
\> $q \gets \op{majority-stable}(V)$ \\
\> $\var{blob} \gets \op{auth-encrypt}((s, V, k_{\CC}), k_{seal})$ \\
\> $\op{store}(\var{blob})$ \\
\> $\var{reply} \gets \op{auth-encrypt}(\msg{reply}{t, h, r, q, h_c}, k_{\CC})$ \\
\> send message $\var{reply}$ to $C_i$
\end{tabbing}
}
\caption{\pp Protocol for trusted execution context $T$}
\label{alg:enclave}
\end{algo}

\subsection{Bootstrapping}

Bootstrapping sets up the necessary cryptographic keys and security
contexts for trusted execution.  It consists of three phases:
(1) creating a trusted execution context~$T$ on a remote server; (2) remote
attestation and provisioning of~$T$; and (3) key distribution among the
group of clients.

In the first phase, a special \emph{admin} client instructs the server to
create a new trusted execution context~$T$ for running protocol~\pp
(Alg.~\ref{alg:enclave}). When~$T$ starts this protocol, it enters 
$\op{init}$ first.
Function \op{init} is also executed after a reboot, where it first
\op{loads} the encrypted state from stable storage.  During initialization
no such state exists yet.

Second, the admin initiates the remote attestation process, to verify
that~$T$ has been started correctly and is running \pp.  Remote attestation
is a core function of the TEE and produces a cryptographic proof, which
convinces the admin that $T$ indeed runs \pp.  If a malicious $S$ would
instantiate~$T$ with some~$P \neq \pp$ this verification will reveal it.
Note that the remote attestation protocol also convinces the admin that~$T$
is actually executed on the TEE and protected against a malicious server.

Finally, the admin generates two secret keys, $k_{\CC}$ for securing the
communication and $k_{P}$ for storing the protocol state, and injects them
into~$T$ through a secure channel provided by the TEE.
After $T$ has received the keys, it initializes the protocol and service
states, and retrieves a sealing key~$k_S = \op{get-key}_{T,\pp}$ from the
TEE.  Recall that $k_P$ is used to encrypt the state, and that $k_P$ and
$k_{\CC}$ together are \op{stored} encrypted~$k_S$.  Since $k_S$ is
generated in a deterministic way in the trusted hardware of the TEE, $T$
can recover its state from an earlier epoch using the stable storage of~$S$
after a crash.  And because every $T$ running on a different physical TEE
obtains a different sealing key, this binds the state of $T$ to the
hardware.
The admin also distributes the communication key to the
clients using a secure channel to each of them.

\subsection{System reboot and recovery}

The server~$S$ controls starting and stopping~$T$.  As argued before, the
TEE is stateless and, therefore, $T$ cannot distinguish a reboot after a
crash from an attack by~$S$.  In order to tolerate server crashes and
reboots without administrative intervention, but also to facilitate planned
restarts, the application state is \op{stored} on stable storage.

When the server reboots after a crash, it recreates the trusted execution
context~$T$ that runs \pp.  $T$ then enters \op{init}, which first tries
to \op{load} a previous state and resumes from there when it
exists.  As $T$ obtains $k_S = \op{get-key}_{T,\pp}$ from the TEE, it can
decrypt and authenticate $k_P$ and the state with~$k_S$; the state also
contains $k_S$ for communicating with the clients.  $T$ recovers $V$ form
the state and can easily derive $(t, h)$ from $V$ by looking up the client
which executed the last operation in~$V$.  Formally, $V$ is an array of
$(t_a, t, h)$ triples, and $\op{argmax}(V)$ returns the index of the triples
with the highest sequence number~$t$.

$T$ has now entered a new epoch and is ready to continue request processing
without remote attestation.  The clients trust that $T$ runs \pp from the
initial verification step during bootstrapping and from the binding of the
sealing key ($k_S$) to the TEE through the secure hardware.  Recall that
$T$ recovers the communication key~$k_\CC$ via the sealing key.  Once a
client can engage in encrypted and properly authenticated communication,
protected through $k_\CC$, to some TEE, the trust of the client from the
initial attestation extends to the current holder of~$k_\CC$.

\subsection{Stability}

For determining the stability of operations, $T$ maintains the map~$V$ with
two sequence numbers for every client. One sequence number of the last
acknowledged operation, and another sequence number of the last operation.
The function $\op{majority-stable}(V)$ returns the sequence number of the
operation that is stable among a majority, that is, the largest acknowledged
sequence number in $V$ that is less than or equal to \emph{more}
than $n/2$ sequence numbers in~$V$.
Stability indicates to the clients when their operations have been observed
by others and helps detecting forking attacks.  When the server is correct
and all clients periodically invoke operations, then all operations become
stable eventually.  In the case of a forking attack, where one or more
clients are separated, the operations of the forked clients will cease
to become stable.

The client protocol returns the sequence number $t$ and the majority-stable
sequence number~$q$ together with the operation result.  This enables the
client to track the progress of the operation history.  Depending on the
application, a client might want to verify that some critical operation has
become stable or wait until it does before invoking new operations.  Note
that the client protocol as described in Alg.~\ref{alg:client} only
receives stability updates when it invokes new operations.  If the client
needs to be informed earlier about the stability of past operations, it can
simply invoke \emph{dummy operations} periodically, as introduced by
FAUST~\cite{cakesh11}.
Alternatively, Alg.~\ref{alg:client} could be extended to support a
callback mechanism, where clients can register for notifications of
stability updates, as also used in Venus~\cite{scckms10}.

\subsection{Extensions}

\subsubsection{Tolerating server crashes}

As the server might crash, we now extend the protocol to allow~$T$ to recover and continue processing.  In the simple case
where~$T$ crashes while it is idling, the correct server restarts it and
continues with the protocol as described before.  On the other hand, when
$T$ crashes during the processing of a client request, we differentiate
between two cases: either it crashes \emph{before} the \op{store} operation
returns and has saved the application and protocol state or
\emph{afterwards}.
 
Therefore, we equip the client with a retry mechanism: When the client has not
received a reply until a timer expires, it sends the message again,
but marks it as a retry attempt.  In the first case ($T$ crashes before
successfully stores), the server will restart $T$ and it
eventually receives the retry message.  The verification of the sequence
number~$t_c$ and the hash chain value~$h_c$ ensures that the lost message
has not already been processed.  $T$ simply continues processing and
returns the reply to the client.  In the second case (when~$T$ crashes
after stores), the verification of $t_c$ and $h_c$ fails since
$t_i$ stored in $V[i]$ is bigger than the value received from~$C_i$.  The
retry marker instructs $T$ to not consider this as a rollback attack.
Therefore, we extend the protocol state~$V$ to store the last operation
result~$r$ as well.  Then $T$ can retrieve the result from~$V$ and
(re)send the \str{reply} message.

\subsubsection{Server migration}

Since location transparency is a major advantage in cloud computing, we
also include a migration mechanisms that allows to move a trusted execution
context~$T$ to a different host system.
There are two trusted execution context instances involved, $T$ on the origin
system and $T'$ on the target system to which the protocol migrates.
Migration requires cooperation between the two machines and that the
server's stable storage can be accessed from the origin and the target
system, for instance by using shared remote storage.

The migration works as follows.  The (correct) origin server signals the
target server to start a trusted execution context~$T'$ and to prepare it
for migration.  Normally, $T'$ would try to retrieve a state encryption key
$k_P$ from stable storage but since it was encrypted with the sealing key
of $T$ on the origin system, $T'$ cannot obtain it.  For this reason, $T$
takes over the role of the admin and bootstraps $T'$ according to the
earlier description.  After a successfully remote attestation, $T$ injects the
state encryption key~$k_P$ via a secure channel.  At this point, $T$ stops
processing requests and provides its current state to~$T'$; then $T'$
restores the application and protocol state, resumes executing requests,
and is still able to uphold the guarantees of \pp against rollback and forking
attacks.

This migration mechanism does not require a trusted party and works
completely transparently for the clients.  However, when the origin system
crashes without any possibility to recover, e.g., when the TEE hardware
malfunctions, then an intervention by a trusted admin is required.
In contrast to the solutions based on a TMC mentioned in
Sec.~\ref{subsec:tmc}, this migration mechanism is more robust to server
failures.  In particular, the migration of a TMC always requires an admin
to read the last TMC value from the origin system and to update the TMC on
the target system with the correct counter value.  Clearly, this fails if
the origin system becomes inaccessible.  \pp still allows migration because
the TEE is stateless and because the state is stored on remote storage.
Our proposed migration scheme is similar to~\cite{strackx2015}.

\subsubsection{Group membership}

In a practical system, the group of clients will dynamically change, as
clients may be removed from the collaboration group and new clients may
join.
Although the protocol formulation uses a static client group, it is easy to
extend \pp for handling dynamic changes.  When a new client joins the
group, the admin sends the shared secret~$k_C$ for secure communication
with the trusted execution context to the new client and instructs $T$ to
include the client in the protocol state.  For removing a client, the admin
generates a new fresh communication key~$k'_C$ and distributes it to all
remaining clients.  Then the admin sends a removal request with~$k'_C$ to
$T$, which uses the fresh key afterwards.

\section{Implementation}\label{sec:impl}

The \pp protocol relies on our assumptions as described in
Sec.~\ref{sec:model} and can be implemented with any TEE technolgy such as Intel SGX.

\subsection{Intel SGX}\label{sec:sgxoverview}

\ac{SGX}~\cite{mckeen2013} adds 
hardware enforced security to the Intel CPU architecture.
\ac{SGX} enables applications to execute certain code in a trusted execution
context, also called \emph{enclave}.  Enclaves are isolated and a hardware
enforced mechanism guarantees the confidentiality and the integrity of an
enclave even if the entire system is compromised.
Moreover, the \ac{SGX} platform checks that an application has not been tampered
with when loading code and data at initialization into an enclave.
\ac{SGX} offers an attestation mechanism~\cite{anati2013} for enclaves that allows
to prove to a remote third party that an enclave runs a given application on
an actual \ac{SGX} platform.
For utilizing the system's persistent storage and at the same time preserving
data confidentiality and integrity, \ac{SGX} supports data sealing.  It permits to
unseal data only by the origin enclave or another enclave by the same
enclave developer.
In the \ac{SGX} programming model, applications in an enclave are considered to be
trusted whereas all other applications (even the operating system) are
untrusted.  Typically, those enclave applications are small, hence, it is less
likely to expose vulnerabilities.
Using the \ac{SGX} Software Development Kit~(SDK)~\cite{sgx-sdk-linux, sgx-sdk-win}
enables developers to divide their applications into a trusted component
(enclave) and untrusted component.  The trusted component is signed by the
developer.
For bridging the trust border between enclaves and untrusted components, SGX
provides the Enclave Definition Language~(EDL) that is used by enclave
developers to specify an interface and generate ``gateway'' code comprising
Enclave calls (ecall) and Outside calls (ocall).

\subsubsection{Enclave protection}

\ac{SGX} features two properties that are essential to execute code securely in an
enclave.
First, \ac{SGX} verifies that an enclave is instantiated with the correct
application.
The enclave code contains an Enclave Signature (SIGSTRUCT)
produced by the enclave developer that allows the \ac{SGX} platform to detect
whether the code of the enclave has been tampered with.  In particular,
SIGSTRUCT comprises an enclave measurement (a cryptographic hash that
identifies the code and data), a signature over the measurement, and the enclave developer's public key, that
serves as the identity of the enclave developer.  When the enclave is loaded,
the CPU verifies the signature and calculates the enclave measurement and compares it to the measurement
in SIGSTRUCT; if they match the enclave completes its instantiation
successfully.
Second, \ac{SGX} protects against any access and modification from untrusted
components.
To this end, the enclave resides in an isolated memory area called enclave
page cache (EPC) that can not be accessed from outside an enclave.  This is
enforced through a memory access control mechanism.  The EPC size is limited
to 128 MB, thus, when enclave reaches that limit or a context switch occurs,
pages are moved to DRAM.  A memory encryption engine~\cite{gueron16}
protects pages when swapping between EPC and DRAM in terms of confidentially,
integrity, as wells provides replay attack prevention.
Those two mechanisms prevent any untrusted component from accessing or
modifying the enclave memory.
Note that this mechanism only protects the in-memory state but not persistent
state of an enclave.  When an enclave is terminated, all in-memory state is
lost.

\subsubsection{Enclave attestation}

\ac{SGX} supports remote attestation~\cite{anati2013} that demonstrates to a remote
client that an enclave runs a given application inside a \ac{SGX} platform
and therefore can be considered to be trustworthy.
This is vital for establishing trust in an enclave application and is required
prior provisioning any secrets or protected data.
The remote attestation briefly works as follows:
A remote client sends a challenge to the enclave including a nonce.  The
enclave produces a report that comprises some metadata including a short hash
value of the application code, the enclave developer identity, and additional
user data. The user data contains the nonce.  Note that enclave developers may
also include custom values in the user data, for instance, some information
about the current enclave state.  Additionally, the report comprises a MAC
that is produced using a report key provided by the \ac{SGX} platform.
A special enclave, so called Quoting enclave, receives the report and
validates it by using the same report key.  The \ac{SGX} platform enforces that
only enclaves are able to retrieve this report key, thus, are able to create
and verify report structures.
If the verification succeeds, the Quoting enclave signs the report with a
platform specific key and replaces the MAC with the signature.  \ac{SGX} leverages
a group signature scheme (EPID~\cite{brickell2011enhanced}) that does not
reveal the identity of the platform.  In other words, the signature states
that some \ac{SGX} platform has produces that signature.  The signed report
(Quote) is sent to the remote client which then validates
signature (using an EPID infrastructure), verifies integrity of the attest, and
finally checks that that the Quote matches the challenge using the nonce.

\subsubsection{Data sealing}

Application code and data are secured while residing within an enclave.
However, when an enclave is terminated the data is lost and can not be
recovered when the enclave restarts again.  Therefore, \ac{SGX} features a sealing
mechanism~\cite{sgx-sdk-linux,sgx-sdk-win} based on
AES-GCM-128 that allows to encrypt and authenticate data before it leaves an
enclave by using a special sealing key provided by the \ac{SGX} platform.
In particular, \ac{SGX} provides two types of sealing: An enclave identity based
sealing that only allows enclaves running the same application to unseal the data; 
and enclave developer based sealing where all enclaves, which are developed (signed) by the same developer, can unseal the data.

\subsection{\pp framework}

We implemented \pp as a framework in Java and C++ consisting of a
client-side and a server-side library that can be integrated with SGX-enabled
applications which require rollback and forking detection for persistent state.
%
%
%
The LCM client-library is implemented in Java and follows the description as
presented in Alg.~\ref{alg:client}.  It uses AES-GCM with 128-bit keys
provided by the Java Cryptography Extension to protect the confidentiality and
integrity of all protocol messages.  The LCM client-library uses a simple
network interface including methods for sending and receiving protocol
message. This allows to reuse an existing application network stack instead of
handling the communication with the server by our library.
%
%
The LCM server-side library is implemented in C++ using the Intel \ac{SGX} SDK
(Version~1.6)~\cite{sgx-sdk-linux}.  It only utilizes trusted libraries
provided by the \ac{SGX} SDK, such as \emph{libsgx\_tcrypto} for cryptographic
hashing and encryption.  In particular, we use SHA-256 for constructing the
hash chain and AES-GCM with 128-bit keys for encrypting the protocol messages,
as well as the protocol and application state. The state encryption key is
encrypted using the \ac{SGX} sealing function before storing persistently.
We defined two interfaces that must be implemented by the enclave application.
First, an operation processor, that receives a client operation and returns the
operation result; and second, a serialization interface that returns the
application state as a byte sequence.
The implementation does not strictly follow the Alg.~\ref{alg:enclave} as
presented in Sec.~\ref{sec:arch}. That is, we optimized the code in order
to eliminate the ocall when storing the application and protocol state at
server's persistent storage.  Instead, we piggyback the encrypted data
together with the reply message.
Furthermore, we implemented operation batching mechanism where the LCM
protocol receives multiple invoke messages with a single ecall.  In contrast
to Alg.~\ref{alg:enclave}, the application and protocol state is stored
once per batch.
Our current proof of concept does not make use of remote attestation. 
However, this can be easily extended using the mechanisms as
provided by the \ac{SGX} SDK.

\subsection{Building applications with LCM}
\label{impl:ss:aps}

\begin{figure}[ht]
    \centering
   \includegraphics[width=12cm]{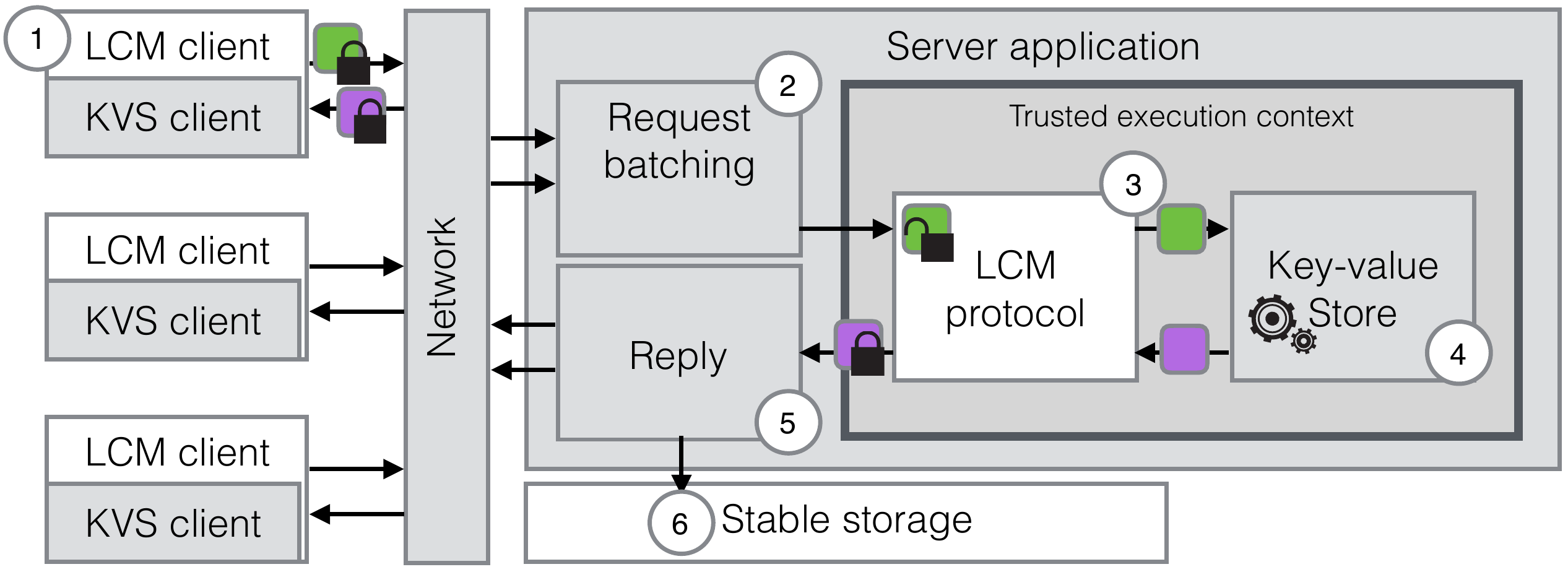}
    \caption{
    The prototype architecture of an enclave based key-value store protected
    with LCM.}
    \label{fig:lcm-proto}
\end{figure}

In order to demonstrate our LCM framework we integrated LCM with a simple
persistent key-value store (KVS) running in an enclave on a remote server. The
prototype architecture is shown in Fig.~\ref{fig:lcm-proto}. Clients and the
server communicating via TCP socket connections.
A KVS stores data objects in a flat namespace, where each object is identified
by a unique name or key.  The KVS is implemented using trusted libraries
provided by the \ac{SGX} SDK. In particular, we use \emph{std::map} for storing
key-values pairs as strings of arbitrary length.  The current version of the
\ac{SGX} SDK does not support \emph{std::unordered\_map} which would be our first
choice due to its constant access time.

Clients invoke \str{get}, \str{put} and \str{del} operations through the KVS
client which instantiates the \pp client-library.  
A server application handles the socket communications, implements an
interface for stable storage (disk), and hosts an enclave running the server-side
LCM protocol and the KVS.
When the server application receives a client request ($\str{invoke}$
message), it collects the message in a bounded queue (batch).  Once the queue
has reached the limit or no more client request are available, the server
application performs an ecall and passes the collected (batched) messages to
the enclave.  The \pp protocol processes them sequentially and returns the
corresponding $\str{reply}$ messages for each client request and the
aggregated application and protocol state.  The server application then,
writes the encrypted states to disk and forwards the $\str{reply}$ messages to
the clients.  Note in order to achieve crash tolerance, the server application
has to write the state synchronously to disk (fsync), this clearly decreases
the performance.
Our prototype implementation of a \pp-protected key-value store comprises
about 4000 sloc, where enclave components comprise around 2200 sloc.  The rest
is for the untrusted server implementation including the storage and network
code.  The KVS client and the \pp client-library add additional 1600 sloc to
the prototype.

\section{Evaluation}\label{sec:eval}

We evaluated the overhead of \pp with a set of benchmarks using YCSB and
compare it against a \ac{SGX}-secured key-value store without
rollback and forking protection.  Furthermore, we compare the performance of \pp
against a trusted monotonic counter approach and unprotected Redis.

\subsection{Experiment setup}

The experiments use a Dell Optiplex 7040 desktop machine with an i7-6700
Intel CPU that is SGX-capable to run the server.  It is equipped with 8~GB of
memory, 1~Gbps network connection and a SSD drive. 
We simulate clients on a virtual machine with 24 virtual CPUs and 8~GB of
memory, running YCSB as workload generator using Oracle Java (JRE 8, build
1.8.0\_111-b14).
All machines run Ubuntu Linux 14.04.4 Server with the generic 4.4.0-47 Linux
kernel.
The evaluation is driven by \emph{YCSB}~\cite{ycsb}, an extensible tool for
benchmarking key-value stores. It supports many different key-value stores,
such as Redis~(\url{http://redis.io}),
Cassandra~(\url{https://cassandra.apache.org/}) and many more. YCSB
comes with a set of core workloads spanning different application scenarios.
For the evaluation we use workload A with a mix of 50/50 \str{put} and
\str{get} operations and show the overall throughput of all clients.  Every
reported data point is taken over a period of 30 seconds.
We integrated the KVS client including the \pp client-library with YCSB.
As a baseline for our experiments we use our KVS (see Sec.~\ref{impl:ss:aps}) 
protected with \ac{SGX}.
For the comparison with Redis and our native KVS implementation we use
Stunnel~(\url{https://www.stunnel.org}) to secure the communication
with the clients.  Redis has been originally designed for deployment in
private networks, thus, it does not support TLS connections.
\pp and the \ac{SGX}-based KVS prototype establish a secure communication with the
clients by using AES-GCM encryption with 128 bit keys.  In order to simplify
the evaluation process we use predefined encryption keys.

\subsection{Enclave memory}

In a preliminary experiment we evaluated the memory consumption of the \ac{SGX}
key-value store.  We inserted one million objects and measured
the enclave heap allocation using \emph{sgx-gdb}.  Each object with a key size
of 40~byte and~100~byte values.
For 300000 objects we measured an allocation of~93~MB enclave memory whereas
we expected only about~40~MB.  It turned out that the KVS implementation based
on \emph{std::map<std::string, std::string>} comes with a memory overhead of
about 134\%.  In particular, the string key-value pairs consume about 280 byte
whereas the map data structure allocates additional~48~byte for each object
for maintaining an internal search structure. 
Moreover, we measured the latency of \str{put} and \str{get} operations for
different number of objects.  As the EPC is limited, we expected a performance
drop when the number of objects increases and the \ac{SGX} driver starts swapping
EPC pages as also reported in~\cite{scone-osdi16,mw16seckeeper}. We observed
that the latency increases up to 240\% when the KVS holds more than 300000
objects.
We refrain from showing the graph due to page reasons.
We assume that this hardware restriction will be addressed in future CPU releases 
and thus choose our further evaluation workloads to fit into the EPC.

\subsection{\pp protocol message}

\begin{figure}[t!]
    \centering
   \includegraphics[width=12cm]{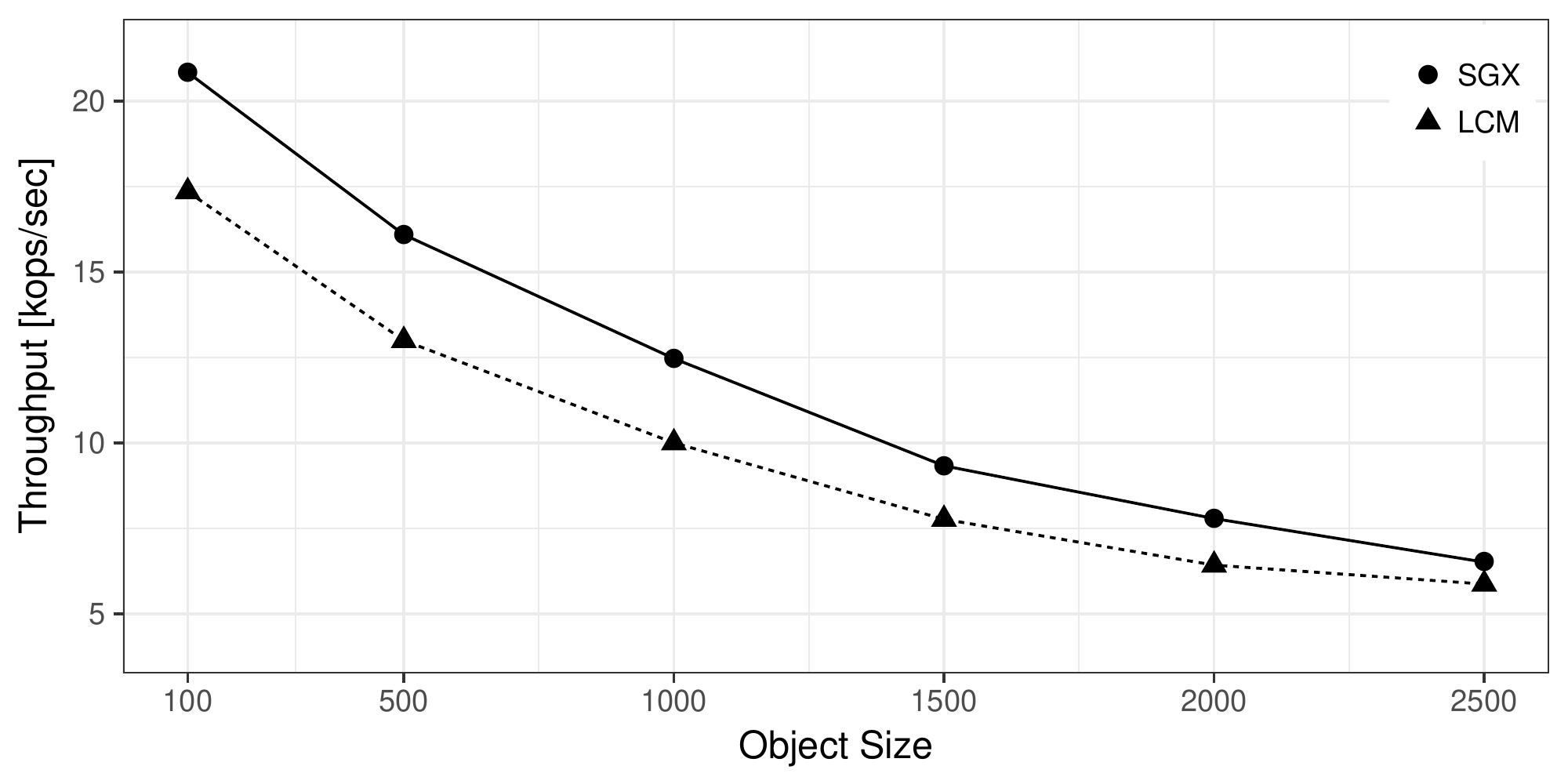}
    \vspace*{-4mm}
    \caption{Throughput with different object sizes with async disk writes.}
    \label{fig:eval:micro-size}
    \vspace*{-4mm}
\end{figure}

We first study how the LCM protocol message overhead affects the throughput.
As described in Sec.~\ref{sec:proto-details}, LCM sends additional
metadata, such as the sequence number and hash chain value, along
with a client request.  In particular, our LCM implementation adds 45 byte to
an operation invocation and 46 byte to a result.  This overhead remains
constant for varying operation and result sizes.
In order to evaluate this, we run the experiment with 8 clients for 1000
objects of size 100 to 2500~byte. Fig.~\ref{fig:eval:micro-size} shows that
the throughput of \pp behaves similar to the plain \ac{SGX} KVS.  As expected, we
observe that \pp introduces an overhead but it decreases with bigger object
sizes. In particular, for objects with the size of 100 byte the throughput is
20.12\% and for objects with size of 2500 byte it is 10.96\%  lower compared
to the plain \ac{SGX} KVS.

\subsection{The throughput of \pp}
\label{sec:eval:tp}

\begin{figure}[ht]
    \centering
   \includegraphics[width=12cm]{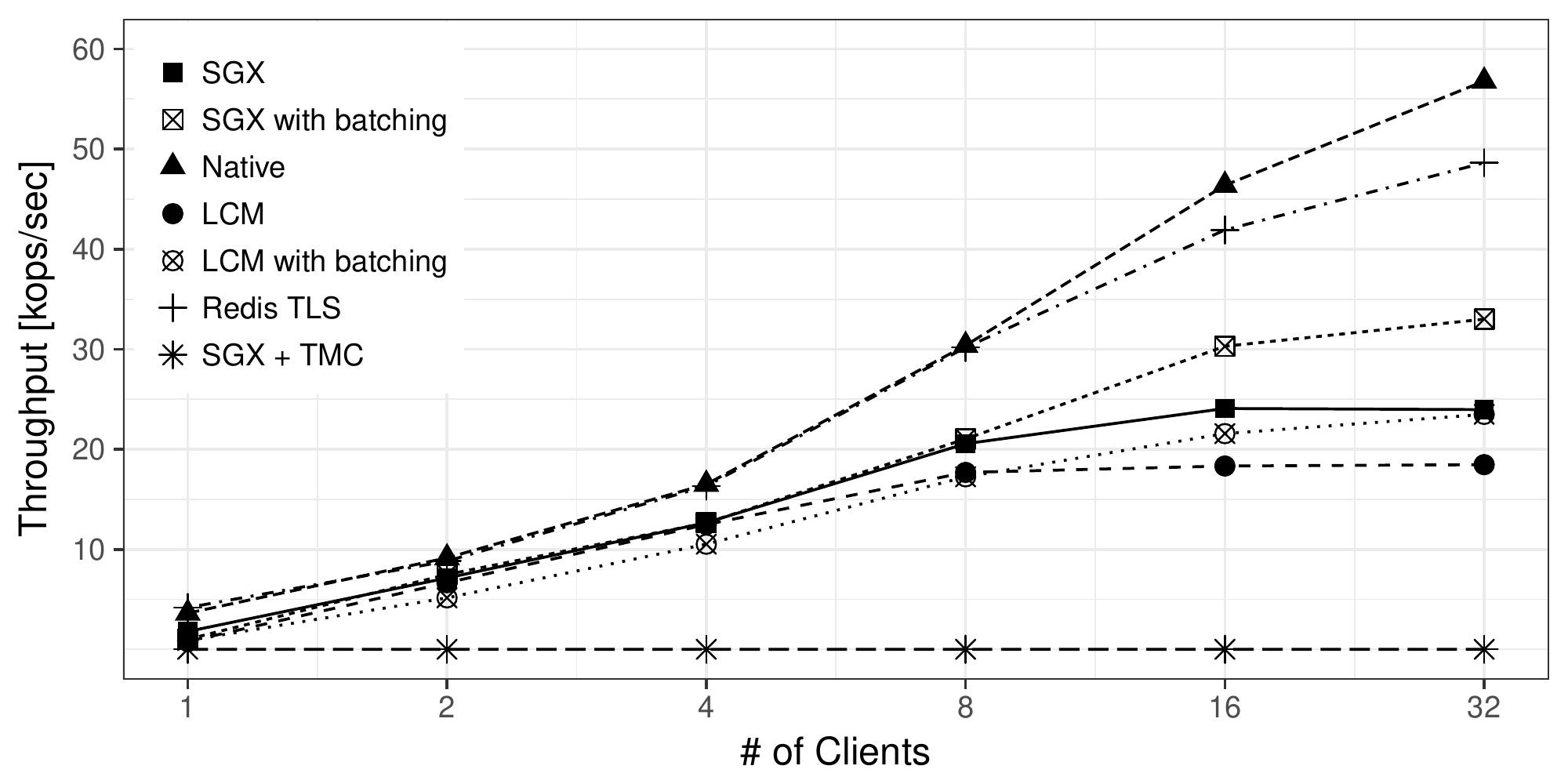}
    \caption{Throughput with different numbers of clients with async disk writes.}
    \label{fig:eval:micro-clients-async}
\end{figure}

We also study the overall throughput of \pp by increasing the number of
clients.  This workload uses up to 32 clients, 1000 objects with a fixed size
of 100 byte.  Each object key is 40 byte.  In this experiment we compare two
variants of \pp against a KVS without \ac{SGX} (``Native''), SGX-secured KVS
(``SGX''), Redis, and SGX-secured KVS with emulated trusted monotonic counter
(``SGX+TMC''). We run \pp and \ac{SGX} without batching enabled and with batching
of up to~16 operations.
We configured Redis to use an append log strategy for persistence. We also
disabled fsync (synchronous disk writes) for Redis as well as for our KVS
prototypes.
As Fig.~\ref{fig:eval:micro-clients-async} shows, the throughput of Redis
and the Native KVS scale almost linear.  In contrast, LCM and \ac{SGX} reach
saturation already with 8 clients.  We observed that the \ac{SGX} KVS reaches 0.42x
-- 0.78x the throughput of the Native KVS.  LCM, on the other hand, reaches
0.67x -- 0.95x the throughput of the \ac{SGX} KVS, with batching even 0.72x -- 0.98x.
The reason is, LCM and \ac{SGX} are single threaded applications and perform the
encryption of every client request inside the enclave. Although, Redis and
Native KVS are also single threaded, they leverage Stunnel that uses multiple
processes to encrypt/decrypt client communication.  That way, secure
communication becomes a bottleneck.

\subsection{The performance impact of Trusted Monotonic Counter}

In this experiment we investigate the performance impact of Trusted Monotonic
Counters (TMC) when used to protect against rollback and forking attacks. The
current version (Version 1.6) of the \ac{SGX} driver and SDK do not yet support
Intel's Trusted
Monotonic Counter~\cite{sgx-sdk-linux} on Linux.  However, on Windows~\cite
{sgx-sdk-win} they are available provided by the Intel management engine (ME)
that stores the counter in non-volatile memory.
We measured an average latency of 60ms to increment a \ac{SGX} TMC on Windows,
whereas~\cite{strpie16} reported even higher latency of about 95ms. We
emulated the TMC on Linux by using a simple counter followed by setting the
thread to sleep for 60ms when incrementing the counter.
As Fig.~\ref{fig:eval:micro-clients-async} shows, the throughput remains
constant for the emulated TMC with an average of 12 operations per seconds,
wheres LCM with batching, on the other hand is 96x -- 2063x faster.  
However, by using trusted monotonic counters rollback
and forking attacks can be detected immediately but this comes with low
throughput.

\subsection{The costs of crash tolerance}

\begin{figure}[ht]
    \centering
   \includegraphics[width=12cm]{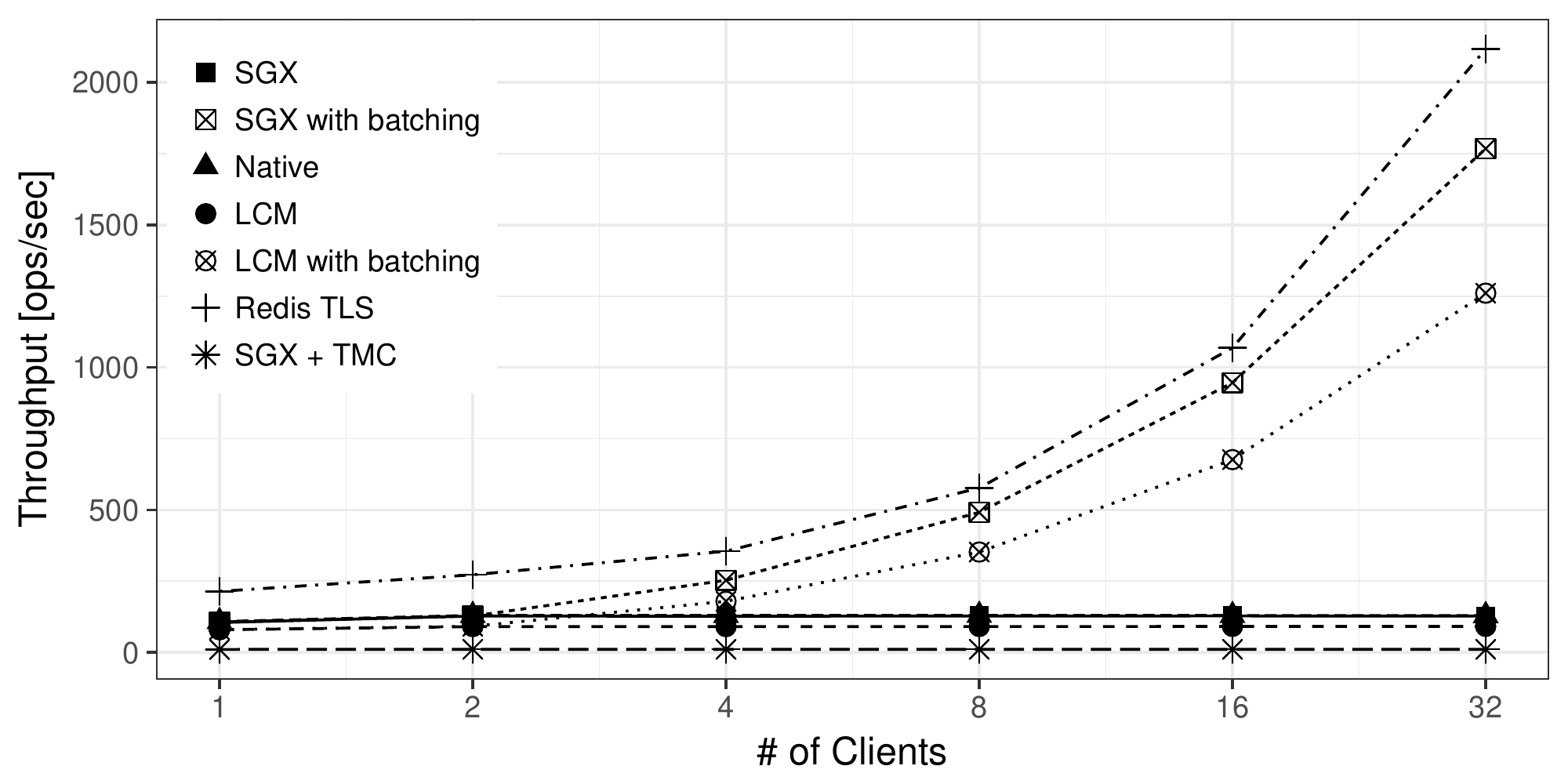}
    \caption{Throughput with different numbers of clients with sync disk writes.}
    \label{fig:eval:micro-clients-sync}
\end{figure}

Finally, we study the performance overhead induced by synchronous disk writes
when storing the application and protocol state that is necessary to support
crash tolerance.
We performed the same experiment as in Sec.~\ref{sec:eval:tp} but enabled
fsync for our KVS prototypes as well as for Redis.  We expect much lower
performance compared to asynchronous writes.
Fig.~\ref{fig:eval:micro-clients-sync} shows the throughput with different
number of clients with synchronous storing.  As expected, fsync introduces high
latency when writing to disk. In particular, we observed that throughput of
Native, \ac{SGX}, LCM, and \ac{SGX} with TMC remain constant whereas Redis, \ac{SGX} and
LCM with batching scale.  \ac{SGX} KVS achieves 0.98x of the Native KVS throughput
and LCM without batching achieves 0.69x of \ac{SGX} KVS throughput.  In contrast,
LCM with batching reaches 0.72x -- 9.87x the throughput of the \ac{SGX} KVS and
0.71x -- 0.75x the throughput of \ac{SGX} KVS with batching.
The experiment shows, that expensive storing operations reduce the relative
overhead introduced by \ac{SGX} but can be reduced by leveraging batching
mechanisms.

\section{Related Work}\label{sec:related}

With the advent of \ac{SGX}, trusted computing has achieved a new level of 
practicality with the aim of wide-spread deployment.
Recent publications detail how legacy applications~\cite{baumann2014},
micro services~\cite{scone-osdi16}, data intensive programming~\cite{vc3} 
but also specific services~\cite{mw16seckeeper} can be secured on top of an infrastructure where \emph{only} the CPU needs to be trusted.
While \ac{SGX} provides special means to detect memory replay attacks against the enclave~\cite{gueron16}, external memory remains unprotected.
Accordingly, additional measures are necessary to prevent rollback and forking attacks mounted through external memory and secondary storage. 
The latter is especially complicated if an enclave is restarted (e.g. due to crashes or system maintenance reboots).
As a pragmatic solution, the Windows \ac{SGX} SDK~\cite{sgx-sdk-win} offers trusted counters that are linked non-volatile memory inside the Intel management engine (ME).
However, trusted non-volatile counters as provided by a TPM are slow, e.g. adding 35-95~ms latency for each operation  depending on the hardware platform, as different reports show~\cite{strpie16,levin2009,rote}.
Thus, in essence, all hardware-based solutions that rely on trusted counters and are consulted on every request of a secured service suffer from performance problems~\cite{chun2007,levin2009,parno2011}. 
An additional issue of current trusted counter TPM-based realizations is wear out if used very frequently.
Strackx and Piessens~\cite{strpie16} specifically address this problem by clever usage strategies, however the performance problems remain.
Recent work~\cite{rote} proposes a complementary approach to \pp where
enclaves across multiple systems assist each other in order to prevent
rollback attacks.  This requires multiple enclaves to interact with each other
to store and retrieve version information from the group of enclaves wheres in
\pp it is stored at the clients.

Another line of work addresses the problem of rollback and forking attacks
without relying on trusted components.  
With only a single client, the classic approach~\cite{blum1994} for memory
checking uses a hash tree where the client stores the root.  Many systems
build on this approach to protect remote storage services (e.g.,
Athos~\cite{Goodrich:2008:AEA:1432478.1432486}). 
In the multi-client model, Mazi{\`e}res et al.~\cite{mazsha02} introduced the
notion of fork-linearizability and implemented SUNDR~\cite{lkms04}, which
confines rollback attacks to always present a view to each client that is
consistent with its past operations; thereby fork-linearizability makes it
much simpler to detect integrity and consistency violations on remote file
storage.
Cachin et al.~\cite{cashsh07} improved the efficiency of SUNDR and proved that
there is no wait-free emulation of fork-linearizable storage. That is,
sometimes clients are blocked until an operation by another client has
finished.  Systems such as SPORC~\cite{fzff10}, FAUST~\cite{cakesh11}, and
Venus~\cite{scckms10} avoid blocking by weakening the consistency guarantees.
Others have explored aborting operations~\cite{mdss09,cacohr14} and improved
the efficiency by reducing the computation and communication
overhead~\cite{brcakn15}.
Mobius~\cite{mobius12} uses forking properties in the context of
disconnected operations.
Previous systems have explored the guarantees of fork-linearizable for different
applications~\cite{fzff10,wisish09} and generic services~\cite{cacohr14}.

\pp combines the best features of these two technologies, trusted execution
environments and protocol-enforced consistency.  It also addresses rollback
and forking attacks on TEEs as much as possible without
introducing impractical limitations into a service.

\section{Conclusion}\label{sec:conclusion}

This work has focused on a shortcoming of trusted computing technology,
which affects current trusted execution environments (TEEs), such as Intel
SGX.  In particular, the trusted execution contexts or ``enclaves'' are
stateless, lose their memory when a crash occurs, and need support from the
host for state continuity.  But since the host is also the adversary of the
TEE according to the security model, it is actually impossible to implement
protocols that survive crashes seamlessly and prevent rollback attacks at
the same time without introducing extra
hardware.
 
As a solution we have introduced \pproj, a system for \emph{detecting}
rollback and forking attacks that ensures the consistency notion of 
fork-linearizable and determines when operations become stable.  The \pp
protocol complements TEE technology with a lightweight mechanism for
maintaining consistency information by the clients.

\section*{Acknowledgments}

We thank 
Anil Kurmus, Cecilia Boschini, Manu Drijvers, Kai Samelin, David Barrera and Raoul Strackx
for interesting discussions and the anonymous reviewers of DSN 2017
for valuable comments.
This work has been supported in part by the European Commission through the
Horizon 2020 Framework Programme (H2020-ICT-2014-1) under grant agreements
number 644371~WITDOM and 644579~ESCUDO-CLOUD and in part by the Swiss State
Secretariat for Education, Research and Innovation (SERI) under contracts
number 15.0098 and 15.0087.

\bibliographystyle{abbrv}
\bibliography{ref}
\balance

\end{document}